# Swirling the weakly bound helium dimer from inside

Maksim Kunitski[1]*, Qingze Guan[2,3], Holger Maschkiwitz[1], Jörg Hahnenbruch[1], Sebastian Eckart[1], Stefan Zeller[1,4], Anton Kalinin[4], Markus Schöffler[1], Lothar Ph. H. Schmidt[1], Till Jahnke[1], Dörte Blume[2,3], Reinhard Dörner[1]*.

[1]Institut für Kernphysik, Goethe-Universität Frankfurt am Main, Max-von-Laue-Straße 1, 60438 Frankfurt am Main, Germany.

[2]Homer L. Dodge Department of Physics and Astronomy, University of Oklahoma, 440 W. Brooks St., Norman, OK 73019, USA.

[3]Center for Quantum Research and Technology, University of Oklahoma, 440 W. Brooks St., Norman, OK 73019, USA.

[4]GSI - Helmholtzzentrum für Schwerionenforschung, 64291 Darmstadt, Germany.

*Correspondence to: kunitski@atom.uni-frankfurt.de, doerner@atom.uni-frankfurt.de.

**Summary paragraph.** Controlling the interactions between atoms with external fields opened up new branches in physics ranging from strongly correlated atomic systems to ideal Bose[1] and Fermi[2] gases and Efimov physics[3,4]. Such control usually prepares samples that are stationary or evolve adiabatically in time. On the other hand, in molecular physics external ultrashort laser fields are employed to create anisotropic potentials that launch ultrafast rotational wave packets and align molecules in free space[5]. Here we combine these two regimes of ultrafast times and low energies. We apply a short laser pulse to the helium dimer, a weakly bound and highly delocalized single bound state quantum system. The laser field locally tunes the interaction between two helium atoms, imparting an angular momentum of $2\hbar$ and evoking an initially confined dissociative wave packet. We record a movie of the density and phase of this wave packet as it evolves from the inside out. At large internuclear distances, where the interaction between the two helium atoms is negligible, the wave packet is essentially free. This work paves the way for future tomography of wave packet dynamics and provides the technique for studying exotic and otherwise hardly accessible quantum systems such as halo and Efimov states.

**Main Text.** In ultracold atomic gas experiments, the two-body interaction is commonly tuned by an external magnetic field in the vicinity of a Feshbach resonance[6,7]. An alternative approach is to use an external electric field [8], which is particularly interesting for weakly bound systems consisting of non-magnetic atoms, where the interaction between field-induced dipoles can significantly alter the native interatomic potential. One such system is the helium dimer, $^4$He$_2$, in which the two helium atoms are held together by an extremely weak van-der-Waals interaction. The corresponding potential is isotropic with a well depth of 11 K (0.95 meV) at an internuclear distance of 3 Å (Fig. 1a and green line in Fig. 1c)[9]. In the absence of external fields, the potential supports a single rovibrational bound state with a tiny binding energy of 1.7 mK (150 neV)[10]. The state's probability distribution is isotropic and spreads far beyond the potential well, so that about 80% of the dimer resides in the classically forbidden tunneling region, justifying the name "quantum halo."

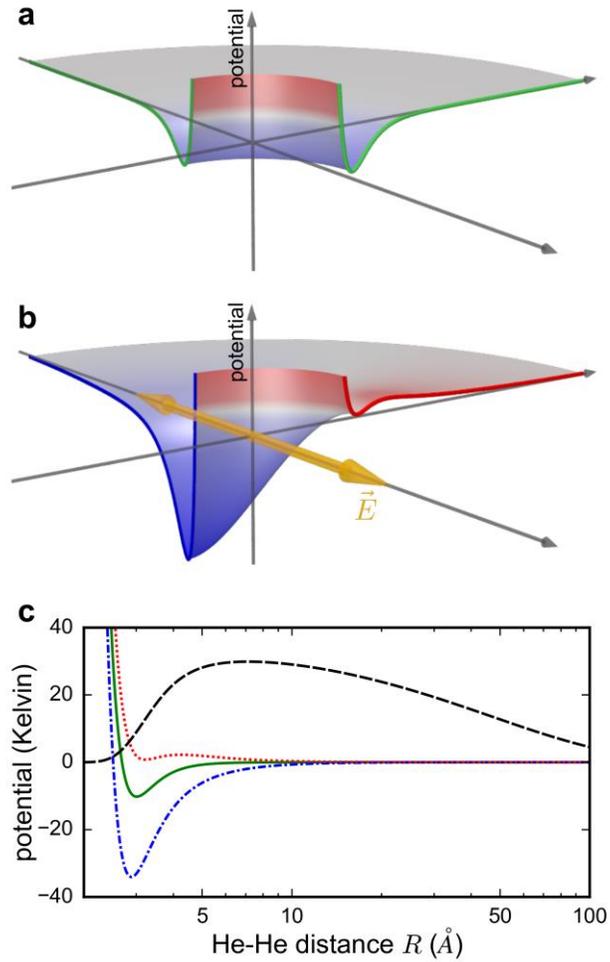

**Fig. 1**. **Field-induced interatomic potential of He$_2$. a**, Native isotropic He-He potential. **b**, Laser-dressed potential for an intensity of $1.3 \cdot 10^{14}$ W/cm$^2$ (E≈$3.13 \cdot 10^8$ V/cm); the field polarization is shown by the orange arrow. **c**, One-dimensional comparison of the native (green solid line) and field-induced He-He potential along the electric field direction (blue dashdotted line) and perpendicular to the field direction (red dotted line). The density distribution corresponding to the native potential is shown by the black dashed line. Note the logarithmic scale on the x-axis.

A foray of theory predictions exists on exciting new phenomena arising from the field modification of the interaction between two helium atoms. They range from tunability of rovibrational states[11–] [14] – see [15–21] for similar effects in alkali molecules – to the emergence of a "covalent"-like dimer that supports many bound states[13,22]. In the three-body sector, the creation of an infinite tower of field-induced $^4$He$^3$He$_2$ Efimov trimers was predicted in the seminal work by Nielsen *et al.*[23]. Experimentally however, these phenomena are unexplored due to challenges such as the realization of high field strengths and preparation and efficient detection of such fragile quantum states. In particular, the following open questions have motivated the current work. How does the quantum halo respond to the non-adiabatic and anisotropic tuning of the interatomic interaction? Does it react to the field as a whole and start to rotate, as a common rigid molecule would? Does one observe an angular anisotropy, even though the native He-He potential does not support bound rotational states?

We applied a strong laser field to the helium dimer and made a movie of the dimer's response using time-resolved Coulomb explosion imaging. We found that the electric field launches a dissociative wave packet at small internuclear distances by locally modifying the He-He interaction potential (Fig. 1b,c). The wave packet propagates towards larger internuclear distances, where it becomes essentially free, resulting in a time-dependent alignment signal (angular anisotropy). Interference of this free wave packet with the residual isotropic ground state wave function allows for the observation of the quantum phase of the wave packet.

Helium dimers in a mass selected supersonic cluster beam[24] are irradiated with a 310 fs 780 nm pump laser pulse with a volume averaged intensity of $1.3 \cdot 10^{14}$ W/cm$^2$ (E≈$3.13 \cdot 10^8$ V/cm), which produces an insignificant amount of ionization. Owing to the anisotropic dipole polarizability, this strong laser field introduces an angle dependence into the He-He interaction. The induced potential is repulsive along the perpendicular direction and its depth along the polarization direction is about 3 times deeper than that of the native potential (see Fig. 1b,c). This response to the external electric field distinguishes the He$_2$ notably from common covalently bound rigid molecules [25], where the laser field induces a tiny overall *attractive* anisotropy. The pump pulse is followed by a delayed and much stronger 30 fs probe pulse (ca. $10^{16}$ W/cm$^2$), which ionizes both helium atoms. The two ions are subsequently driven apart by their mutual Coulomb repulsion, resulting in so-called Coulomb explosion, which is imaged by a COLTRIMS reaction microscope[26,27] (see Supplementary Information). This imaging yields the internuclear distance and the orientation of individual dimers at the instant of ionization by the probe pulse.

The response of a diatomic molecule to a laser field can be quantified by the alignment merit, $\cos^2\theta$, where $\theta$ is the angle between the internuclear axis and the polarization direction of the pump laser. Initially, the He$_2$ molecule is in a pure *s*-state with isotropic angular distribution at all internuclear distances. This corresponds to the expectation value of $\cos^2\theta$ of 1/3 (negative delays in Fig. 2a; see also Fig. 2c, middle). About 1 ps after applying the 310-fs laser pulse (pump), we observe a partial alignment along the polarization direction of the pump field at small internuclear distances ($\langle\cos^2\theta\rangle \approx 0.42$, Fig. 2a), while the far-out regions of the wave function are still isotropic. The alignment wave evolves towards larger internuclear distances such that for a delay time of 40 ps, the inner part of the wave function has restored its spherical symmetry while the outer part (30-70 Å) displays R-dependent alignment. The experimental alignment evolution is well reproduced by our parameter-free quantum simulation (Fig. 2b, see Supplementary Information). The corresponding field-induced evolution of angular distributions can be found in the Supplementary Information.

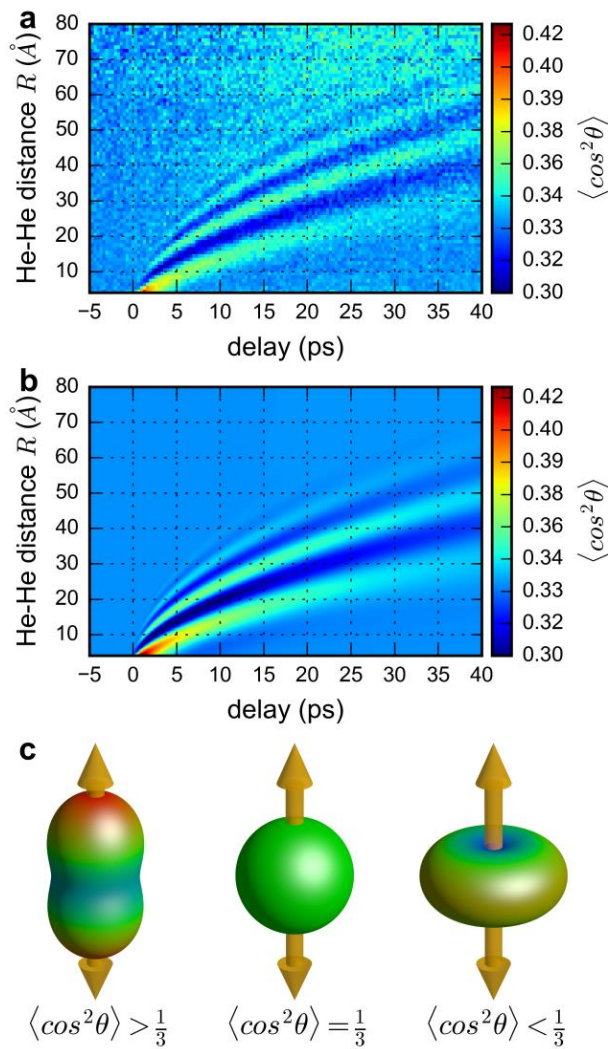

**Fig. 2**. **Temporal evolution of field-induced alignment. a**, Expectation value of $\cos^2\theta$ (color-coded) ($\theta$ is the angle between the helium dimer axis and the pump polarization direction) as function of the internuclear distance and the delay time between the pump and probe pulses. **b**, Quantum calculation for the same parameters as in the experiment. Theoretical probabilities were convolved with Gaussian functions of variable width prior to the calculation of $\cos^2\theta$, emulating the $R$-dependent experimental resolution (see Supplementary Information). **c**, Schematic illustration of the angular distribution of the dimer axis with respect to the laser field polarization (orange arrow) for different expectation values of $\cos^2\theta$.

A key finding revealed by Fig. 2 is that the response of the He$_2$ molecule to the laser field starts in a very narrow small-distance region and then propagates outward to larger internuclear distances. This can be understood from the fact that the modification of the native potential is only significant at small internuclear distances (Fig. 1c), resulting in an initially narrowly confined probability flux (see Supplementary Information) that tends to align the dimer axis along the field. Due to the large extent of the initial state wave function, the helium dimer does not react as a whole to such localized "kick". This is completely different from what is known from all alignment experiments performed to date, where molecules respond to the laser "kick" as a single unit and can therefore be treated within the (semi-) rigid rotor approximation[5]. This response is commonly understood as a coherent excitation of many rotational bound states, *i.e.*, the emergence of a rotational wave packet. The temporal evolution of such a wave packet results in a distance-independent but time-periodic alignment signal. The field-free He-He interaction potential, in contrast, does not support any rotationally excited bound states and the He$_2$ wave packet dynamics, which is characterized by a strong coupling of the rotational and vibrational degrees of freedom, involves the dissociation continuum.

Another way of looking at the field-induced dynamics is to visualize the He$_2$ density. Figs. 3a and 3b color-code the pump laser induced modifications of the probability distribution for $\theta \approx 0°$ and for $\theta \approx 90°$ as functions of the internuclear distance R and the delay time. Remarkably, our measurement allows to resolve the phase of the outward propagating density wave.

In quantum mechanics, the phase of the time-dependent wave packet can only be observed by letting different wave packet portions interfere, which is the basis for any kind of interferometry [28,29]. Since the ground state wave function $\Psi_{GS}(R)$ of the helium dimer extends to large internuclear distances, it provides a real valued, unstructured and isotropic quantum background with which the outgoing dissociative wave packet portion can interfere, thereby unveiling the quantum phase.

Writing the observed time-dependent wave packet $\Psi_{obs}(\vec{R},t)$ as a superposition of partial waves, $\Psi_{obs}(\vec{R},t) = \sum_{J=0,2,...} \psi_J(R,\theta,t)$ with $\psi_J(R,\theta,t) = R^{-1}u_J(R,t)Y_{J0}(\theta)$ ($J$ is the orbital angular momentum quantum number and $Y_{J0}(\theta)$ are spherical harmonics), the density modifications are, to leading order, described by

$$2\pi R^2(|\Psi_{obs}(\vec{R},t)|^2 - |\Psi_{GS}(R)|^2) \approx 4\pi R \Psi_{GS}(R)|u_2(R,t)|Y_{20}(\theta)\cos(\gamma_2(R,t)), \tag{1}$$

where $\gamma_2(R,t)$ is the position- and time-dependent phase of the wave packet (see Supplementary Information). The $\cos(\gamma_2(R,t))$ dependence and, correspondingly, the finger-like structure in Figs. 2a,b and 3a,b are a direct consequence of the fact that the dissociating $J=2$ partial wave component is interfering with the $J=0$ partial wave component (see Supplementary Information for a movie).

Figs. 3c and 3d show that the $\theta=0°$ and $\theta=90°$ waves are out of phase. This is in agreement with Eq. (1), which states that the density modulations at different $\theta$ but fixed $R$ and delay time differ by the value of $Y_{20}$, giving $Y_{20}(\theta=0°) = -2Y_{20}(\theta=90°)$. The fact that the pump laser under our experimental conditions populates predominantly the $J=0$ and $J=2$ partial waves explains why two observables, namely, alignment (Fig. 2) and changes of the probability distribution (Fig. 3), show similar dynamics.

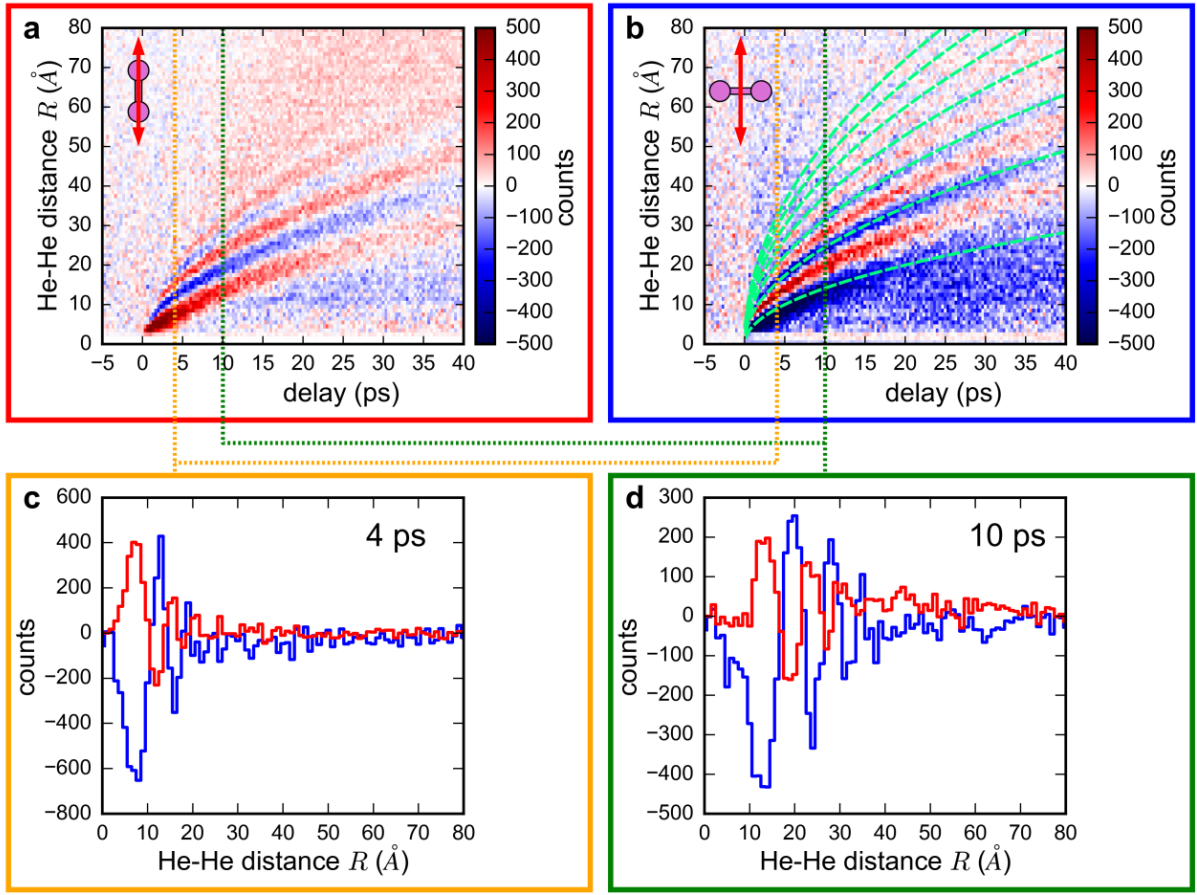

**Fig. 3. Time evolution of the dissociative wave packet in He$_2$ after the laser "kick"**. **a**, **b**, Experimentally determined change of probability $2\pi R^2(|\Psi_{obs}(\vec{R},t)|^2 - |\Psi_{GS}(R)|^2)$ (color coded) as functions of the internuclear distance and the delay time for the cases where the dimer axis is (**a**) parallel to the pump polarization (to make the plot, we consider $\theta = 0 \pm 40°$) and (**b**) perpendicular to the pump polarization (to make the plot, we consider $\theta = 90 \pm 40°$). The green dashed lines in panel **b** show the calculated constant-phase evolution of a free particle with a reduced mass of 2 amu (see Supplementary Information). **c**, **d**, To highlight the difference between the probability change for $\theta = 0°$ and $\theta = 90°$, the red and blue lines show fixed-$R$ cuts through the data shown in panels **a** and **b**, respectively, for two delay times, namely (**c**) 4 ps (orange vertical lines in **a** and **b**) and (**d**) 10 ps (green vertical lines in **a** and **b**).

Since the He-He interaction potential (Fig. 1a) can be neglected at large internuclear distances, the unbound part of the wave packet (Fig. 3a,b) describes a quantum particle moving freely without the influence of any forces. Thus, our experiment directly visualizes the temporal evolution of a propagating free-particle wave packet with the otherwise unobservable quantum phase.

The observed response of a single-state quantum halo, $He_2$, to an intense short electric field can be summarized as follows (Fig. 4). The laser field "grabs on" to a small portion of the ground state wave function of $He_2$ at small internuclear distances and transfers a small fraction of the ground state population to the repulsive $J = 2$ potential, creating a localized dissociative wave packet in the continuum (pump arrow at t=0 ps in Fig. 4). During the subsequent time evolution of the wave packet, the potential energy is converted to kinetic energy. Concurrently, the envelope of the wave packet spreads with time due to intrinsic dispersion of the matter wave. At large internuclear distances, where the interaction potential is negligible, the wave packet evolution resembles that of a free quantum particle with a mean group velocity of about 2 Å/ps (see Supplementary Information). Not only the envelope, but also the phase of this $J = 2$ wave packet is accessible during the measurement due to its interference with the spatially extended isotropic $J = 0$ ground state wave function $\Psi_{GS}(R)$.

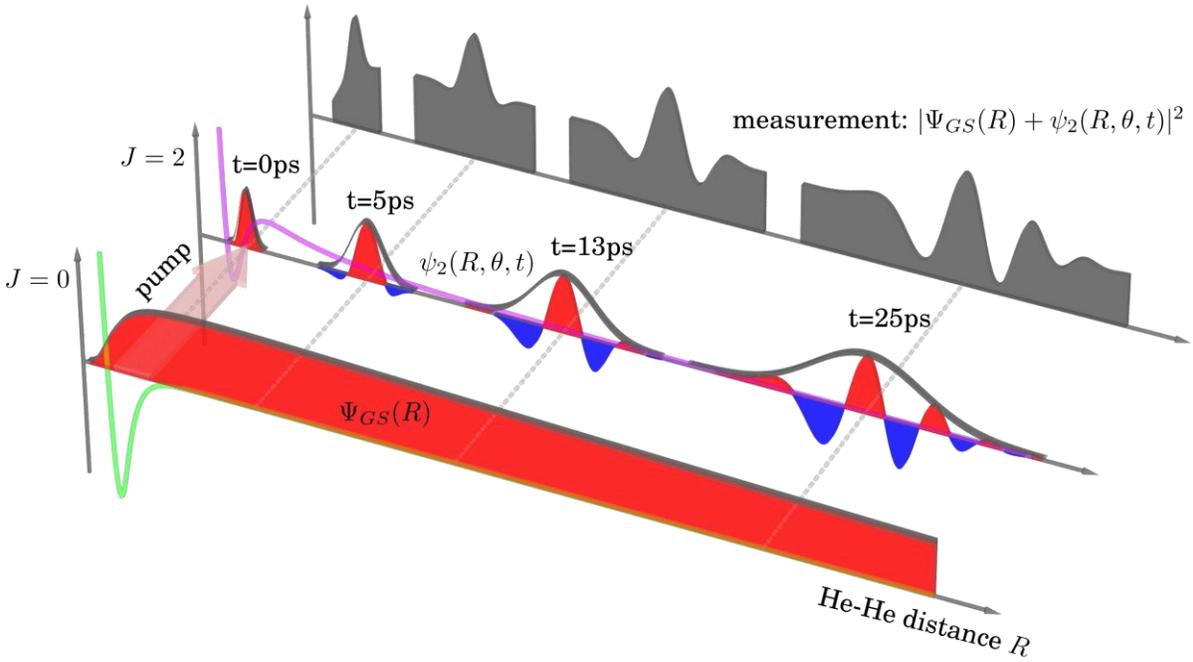

**Fig. 4.** Radial response of the highly delocalized single state helium dimer to a strong laser field (pump). At t=0 ps the pump field transfers a tiny part of the ground state wave function $\Psi_{GS}(R)$ at small internuclear distances to the dissociative $J = 2$ potential. The evolution of this dissociative wave packet consisting mainly of the partial wave $\psi_2(R, \theta, t)$ is imaged by Coulomb explosion imaging. The phase of the wave packet is resolved due to its interference with the spatially extended $J = 0$ ground state wave function $\Psi_{GS}(R)$. Red and blue colors correspond to the positive and negative values, respectively, of the real part of $\Psi_{GS}(R)$ and $\psi_2(R, \theta, t)$. For illustrative purposes, only the radial part of the partial wave $\psi_2(R, \theta, t)$ is shown.

The results reported here provide the first experimental evidence that the $^4$He$_2$ potential can be tuned through the application of an external electric field. Largely delocalized universal quantum systems such as the helium dimer give a powerful twist to wave packet interferometry. These systems have rather simple ground states of a huge spatial extent, which are frequently described by effective low-energy theories. Such initial states can serve as a reference background with which dissociative wave packet portions, launched by a short and intense laser pulse, can interfere. Imaging the corresponding superposition state then directly reveals the temporal evolution of the wave packet's amplitude and quantum phase. This technique, which uniquely combines universal low-energy physics with short high-intensity pulses, can be extended to few-body quantum systems such as the helium trimer, where the field-induced dynamics is completely unexplored. The strong electric field can also be used to create superpositions of *bound* states in quantum systems[30]. An exciting example would be to create a superposition of the ground state and excited Efimov state of $He_3$ and to subsequently watch the birth and decay of an Efimov state in time.

**Data availability:** All experimental data has been archived at the Goethe-University of Frankfurt am Main and is available upon request.

**Acknowledgments:** We are indebted to Bretislav Friedrich for inspiring this work over many years and to Mikhail Lemeshko for many helpful discussions and calculations. The experimental work was supported by Deutsche Forschungsgemeinschaft (DFG). This work used the OU Supercomputing Center for Education and Research (OSCER) at the University of Oklahoma (OU). QG and DB acknowledge support by the National Science Foundation through grant number PHY-1806259.

**Author contributions:** M. K., H. M., J. H., S. E., S. Z., A. K., M. S., L. P. H. S. and T. J. contributed to the experiment. M.K. and R.D. analyzed the experimental data, Q. G. and D. B. developed the theory, Q. G. performed the calculations, all authors contributed to the manuscript.

**Competing interests:** The authors declare no competing interests.

Additional information

**Supplementary information** is available for this paper at…

**Correspondence and requests for materials** should be addressed to M.K.